\documentclass[twocolumn,showpacs,superscriptaddress,amssymb,10pt,pra,floatfix]{revtex4-1}

\usepackage{graphicx}
\usepackage{dcolumn}
\usepackage{bm}
\usepackage{epsfig}
\usepackage{color}
\usepackage{longtable}
\usepackage{hyperref}
\hypersetup{hidelinks}

\newcommand{\p}{\partial} 				
\newcommand{\e}{\mathrm{e}}				
\renewcommand{\c}{\mathrm{c}}			
\renewcommand{\i}{i}					
\renewcommand{\d}{\mathrm{d}}			
\newcommand{\gs}{\mathrm{g}_s} 			
\newcommand{\g}{g}						
\newcommand{\varbeta}{\theta}			
\renewcommand{\vec}{\bm}				
\renewcommand{\S}{\mathcal{S}}			
\renewcommand{\equiv}{=}

\begin{document}

\preprint{}
%
%
%
%
\title{Gravitational effects on geonium and free electron $\gs$-factor\\ measurements in a Penning trap}
%
%
%
%

\author{S.~Ulbricht}
\email{sebastian.ulbricht@ptb.de}
\affiliation{Physikalisch--Technische Bundesanstalt, D--38116 Braunschweig, Germany}
\affiliation{Technische Universit\"at Braunschweig, D--38106 Braunschweig, Germany}

\author{R.~A.~M\"uller}
\affiliation{Physikalisch--Technische Bundesanstalt, D--38116 Braunschweig, Germany}
\affiliation{Technische Universit\"at Braunschweig, D--38106 Braunschweig, Germany}

\author{A.~Surzhykov}
\affiliation{Physikalisch--Technische Bundesanstalt, D--38116 Braunschweig, Germany}
\affiliation{Technische Universit\"at Braunschweig, D--38106 Braunschweig, Germany}

\date{\today \\[0.3cm]}

%
%
%
%

\begin{abstract} 
	We present a theoretical analysis of an electron confined by a Penning trap, also known as \emph{geonium}, that is affected by gravity. 
	In particular, we investigate the gravitational influence on the electron dynamics and the electromagnetic field of the trap. 
	We consider the special case of a homogeneous gravitational field, which is represented by Rindler spacetime. 
	In this spacetime the Hamiltonian of an electron with anomalous magnetic moment is constructed.
	Based on this Hamiltonian and the exact solution to Maxwell equations for the field of a Penning trap in Rindler spacetime, we derive the transition energies of geonium up to the relativistic corrections of $1/\c^2$. 
	These transition energies are used to obtain an extension of the well known $\gs$-factor formula introduced by L.~S.~Brown and G.~Gabrielse [Rev. Mod. Phys. 58, 233 1986].	
\end{abstract}
\maketitle
%
%

\section{Introduction} \label{section:I}
One way to study the properties of a single electron is to analyze the trapped electron in a well known electromagnetic field configuration.
However, extracting characteristics of a free particle from transitions of a trapped one requires a deep understanding of the trapping conditions. 
For this purpose, commonly a Penning trap is used in modern high precision experiments \cite{Gab08,Gab06,Sturm14,BASE}. 
Such a trap weakly confines the particle under usage of an electric quadrupole and a constant magnetic field. For the case of an electron, this leads to bound states with discrete energy levels \cite{Brow86,Dehm88}. 
The transitions in such an \emph{artificial atom}, called \emph{geonium}, are used, for example, to determine the free electron $\gs$-factor. 
This quantity is a dimensionless measure of the electron's magnetic moment $\vec{\mu}$ in the unit of Bohr magnetons $|\vec{\mu}_B|=e\hbar/(2m)$ 
\begin{equation}
\vec{\mu}=\gs \,\vec{\mu}_B\quad.
\end{equation}
While in Dirac's theory \cite{Dirac28} the $\gs$-factor is ${\gs}_{\mathrm{Dirac}}=2$, in practice QED effects lead to deviations from this value.
A few years ago, D.~Hanneke \emph{et al.} have reported high accuracy Penning trap measurements, which determine $\gs=2.002\,319\,304\,361\,46(56)$ \cite{Gab08,Gab06}.
This experimental result is in outstanding accordance with the calculations of T.~Aoyama \emph{et al.} \cite{Aoy18}.
Such an interplay of theory and experiment can help to test fundamental properties of quantum field theory and to search for physics beyond the standard model, see for example \cite{Flam03,Flam08} and references therein.
%

The $\gs$-factor experiments in Penning traps, as they are carried out by \cite{Gab08,Gab06}, are not performed in an isolated environment, but in the presence of the gravitational field of the Earth.
This gravitational field distorts both, the electron dynamics and the electromagnetic field configuration of the trap.
In this contribution, therefore, we perform the theoretical analysis of the effects of gravity on the result of Penning trap experiments. In particular, we also take into account gravitational effects on the electromagnetic field of the Penning trap, which in turn affects the motion of the electron. 
While gravitational effects on the bound electron $\gs$-factor \cite{Jent13} and the cyclotron motion of the electron \cite{Moris15} have been considered, to the best of our knowledge, an analysis of the gravitational influence on Penning trap experiments have not been reported before.\vspace{0.84em}
%

In order to understand, how gravity influences Penning trap experiments, it is natural to describe both, the electron and the electromagnetic field of the Penning trap, in curved spacetime.
In our study, we will consider the case of a homogeneous gravitational field, which is a good approximation for gravity at the surface of the Earth, as we will discuss in Sec.~\ref{section:IIa}. 
The Dirac Hamiltonian, which describes the dynamics of an electron with anomalous magnetic moment in this spacetime, is obtained in Sec.~\ref{section:IIb}.
While this Hamiltonian can be applied for any electron velocities and gravitational field strengths, we aim to use it to describe Penning trap experiments, which are performed in the non-relativistic regime. 
Therefore, in Sec.~\ref{section:IIc} we perform a Foldy-Wouthuysen transformation to obtain the non-relativistic Hamiltonian and its $1/\c^2$-corrections.
Of course, this Hamiltonian accounts not only for gravitational effects, but also for the coupling to the electromagnetic field of a Penning trap.
In Sec.~\ref{section:IIIa} this field is presented as an exact solution to Maxwell equations in the spacetime of homogeneous gravity.
Using first order perturbation theory, we determine the eigenenergies of geonium exposed to gravity up to order $1/\c^2$ in Sec.~\ref{section:IIIb} and Sec.~\ref{section:IIIc}.
Finally, these energies are used to derive an expression for the free electron $\gs$-factor, which generalises the well known results of L.~S.~Brown and G.~Gabrielse.
The summary of our results is given in Sec. \ref{section:IV}.
%
\vfill
\section{Electron in homogeneous gravitational field} \label{section:II}
\subsection{The homogeneous gravitational field\\ in general relativity} \label{section:IIa}
On Earth the biggest empirical effect of gravity is the acceleration of $\g=9.81\, \mathrm{m}/\mathrm{s}^2$ pointing downwards.
In the Newtonian theory of gravity, a vector field of constant acceleration $\vec{\g}$ is a suitable approximation of the gravitational field perceived by this observer. Higher order effects, accounting for the Earth as a spherical body, can be neglected in a small environment of the observers position.
%

The approximation of a homogeneous acceleration $\vec{\g}$  also holds in general relativity in terms of the non-geodesic motion of an observer; Bound to Earth's surface, the observer is not able to follow gravity in a \emph{free fall}. 
In a general relativistic framework, the Newtonian gravitational field of the Earth is replaced by the famous Schwarzschild spacetime \cite{Schwarz16}.
At the surface of the Earth, this spacetime can be approximated by so called Rindler spacetime \cite{Rind60,Rind66}, which is merely flat Minkowski spacetime, but seen by an accelerated observer.
At this level of approximation, there is no spacetime curvature, but a \emph{distortion} of spacetime by acceleration.
%

In order to describe physics perceived by an accelerated observer, we start with the line element $\d s^2$ of Minkowski spacetime and perform a coordinate transformation towards a coordinate system, that describes the reference frame of the accelerated observer.
The Minkowski line element, expressed in terms of Cartesian coordinates $\tilde{\vec{r}}=(\tilde x, \tilde y, \tilde z)$ and proper time $\tau$, reads
\begin{equation} 
\d s^2 =\eta_{\mu\nu}\,d x^\mu \d x^\nu = \d(\c \tau)^2 - \d {\tilde x}^2- \d {\tilde y}^2- \d {\tilde z}^2\quad, \label{eqn:Minkowski}
\end{equation}
where we introduced the metric tensor $(\eta_{\mu\nu})=\mathrm{diag}(1,-1,-1,-1)$.
In this sign-convention time-like distances are described by positive values of the line element $\d s^2>0$.
Moreover, we use Einsteinian sum convention, which means that a sum is performed from $0$ to $4$ when paired Greek letters appear.
The index $0$ is set to be the index of the time-like coordinate.
%

The coordinates of Rindler spacetime are related to the coordinates of Minkowski spacetime by 
\begin{eqnarray}
x'&=&\tilde x \quad, \nonumber\\
y'&=&\tilde y \quad,\nonumber\\
z'&=&-\frac{\g}{2}\,\tau^2+\tilde z \left(1+\frac{\g \tilde z}{2 \c^2}\right) \quad,\label{eqn:RindlerTrafo}\\
\c t&=& \frac{\c^2}{2\g}\,\mathrm{log}\left(\frac{\c^2 +\g (\c \tau+\tilde z)}{\c^2 -\g (\c \tau-\tilde z)}\right)\quad, \nonumber
\end{eqnarray} 
as discussed in \cite{Rind60}.
Here $t$ is the proper time of the observer, located in the center of the accelerated frame. This frame is denoted by the primed coordinates $\vec{r}'=(x',y',z')$. 
%

In order to describe physical processes in the accelerated frame, we use the transformation (\ref{eqn:RindlerTrafo}) to derive the line element of Rindler spacetime
\begin{eqnarray} 
\d s^2  &=& \left(1+\frac{2\g z'}{\c^2}\right)\d(\c t)^2 \label{eqn:firstRindlerMetric}\\
& &\quad - \d {x'}^2- \d {y'}^2- \left(1+\frac{2\g z'}{\c^2}\right)^{-1}\d {z'}^2\quad. \nonumber
\end{eqnarray}
From now on, we will call $t$ the \emph{time} and only perform spatial coordinate transformations, if required by the geometry of the problem under consideration.
For our further investigation, it is useful to introduce the auxiliary coordinate $u(z')=(\sqrt{1+2\g z'/\c^2}-1)\,\c^2/\g$, as described in \cite{Rind66}, such that the spatial part of the line element is isotropic 
\begin{eqnarray} 
\d s^2 &=& \left(1+\frac{\g u}{\c^2}\right)^2\d(\c t)^2 - \d {x'}^2- \d {y'}^2- \d u^2 \label{eqn:RindlerMetric}\\
&=& g_{\mu'\nu'}(u)\,\d x^{\mu'}\d x^{\nu'}\quad, \nonumber
\end{eqnarray}
%

As seen from Eq. (\ref{eqn:RindlerMetric}), the line element in the accelerated frame is coordinate dependent. Therefore, the measure of time is different at different heights and the factor $\left(1+\g u/\c^2\right)$ in front of the infinitesimal time step $\d (\c t)$ gives rise to the gravitational redshift \cite{Rind60,Rind66}.
For vanishing acceleration the Rindler line element (\ref{eqn:RindlerMetric}) reduces to the Minkowski line element.
The same holds in the $(x',y')$-plane, where we reach flat Minkowski spacetime asymptotically for $u\to 0$.
Therefore, it is legitimate to apply methods of quantum mechanics in Minkowski spacetime in a small area around the coordinate center and treat the modifications, caused by deviation from Minkowski spacetime, as corrections later. 
In our case this assumption is valid, since we are interested in quantum objects bound close to $\vec{r}'=0$, where typical length scales $z'$ are in the micrometer domain and, therefore, much smaller than $\c^2/\g\sim \mathrm{ly}$, which is the typical length scale for the considered gravitational effects.  
%

\subsection{Electron with anomalous magnetic moment\\ in Rindler spacetime} \label{section:IIb}
In the previous section, we introduced the spacetime of a homogeneously accelerated observer, known as Rindler spacetime.
Now we want to pay particular attention to the dynamics of an electron in this spacetime.
Again we start our analysis in Minkowski spacetime, where the electron motion is described by the Dirac equation \cite{Dirac28}
\begin{equation}
(\i \hbar \gamma^\mu\p_\mu-m\c)\psi(x^\nu)=0\quad. \label{eqn:Dirac}
\end{equation}
Here $\psi(x^\nu)$ is the Dirac spinor, whose four components represent not only the electron, but also the positron in their two spin states. 
The Dirac matrices $\gamma^\mu$ are chosen such, that their anti-commutator generates the metric tensor of the line element (\ref{eqn:Minkowski}) of Minkowski spacetime
\begin{equation}
\{\gamma_\mu,\gamma_\nu\}=2\,\eta_{\mu \nu}\quad,  \label{eqn:Clifford}
\end{equation}
where the upper index of the Dirac matrices in Eq.~(\ref{eqn:Dirac}) is raised by the inverse metric tensor $\gamma^\mu=\eta^{\mu\nu}\gamma_\nu$.
%

In the following, we transform the Dirac equation (\ref{eqn:Dirac}) to the frame of an accelerated observer.
For this purpose, it is convenient to start with the Dirac action in Minkowski spacetime
\begin{equation}
S[\bar\psi,\psi]=\int\bar\psi(x^\nu)(\i \hbar \gamma^\mu\p_\mu-m\c)\psi(x^\nu)\,\d x^4\quad, \label{eqn:DiracAction}
\end{equation}
where $\bar \psi(x^\nu)=\psi^\dagger(x^\nu) \gamma^{0}$ is the Dirac adjoint of $\psi(x^\nu)$  and $\int \d x^4$ is the integral over all four spacetime coordinates $(x^{\mu})=(\c \tau, \tilde x,\tilde y,\tilde z )$.
%

From action (\ref{eqn:DiracAction}) the Dirac equation (\ref{eqn:Dirac}) can be obtained by the principle of stationary action. 
The next step is to express the Dirac action in the coordinates $(x^{\mu'})=(\c t,  x', y', u)$ of the accelerated frame, in order to obtain Dirac equation in Rindler spacetime.
This coordinate transformation of the action requires some attention \cite{Hehl90,Lippoldt,Weldon} and, therefore, is discussed in Appendix \ref{section:A}.
After the coordinate transformation (\ref{eqn:RindlerTrafo}), the action (\ref{eqn:DiracAction}) reads
\begin{eqnarray}
S[\bar\psi',\psi']=\int\bar\psi'(x^{\nu'})(\i \hbar \gamma^{\mu'}(u)\p_{\mu'}-m\c)\psi'(x^{\nu'}) \qquad\qquad \label{eqn:DiracAction1}& &\\
\times \left(1+\frac{\g u}{\c^2}\right)\,\d {x'}^4\quad, \nonumber & &
\end{eqnarray}
where the infinitesimal spacetime volume $\d {x}^4$ in now replaced by $\left(1+\g u/\c^2\right)\, \d {x'}^4$. 
In addition the primed Dirac matrices in Eq.~(\ref{eqn:DiracAction1}) are spacetime dependent and have to obey the relation
\begin{equation}
\{\gamma_{\mu'}(u),\gamma_{\nu'}(u)\}=2\,\g_{{\mu'} {\nu'}}(u)\quad, \label{eqn:Clifford2}
\end{equation}
instead of the relation (\ref{eqn:Clifford}). Here $\g_{{\mu'} {\nu'}}(u)$ is the metric tensor of Rindler spacetime, defined in Eq.~(\ref{eqn:RindlerMetric}). The Dirac adjoint spinor now reads $\bar \psi'(x^{\nu'})=(\psi'(x^{\nu'}))^\dagger \gamma^{0'}(u)$.
%

Indeed, in the case of Penning trap experiments, the electron is not only exposed to gravity, but is located in an electromagnetic field.
Therefore, we go the common way to introduce a minimal coupling to the electromagnetic field by the replacement of the partial derivative
$\p_{\mu'}\to \p_{\mu'}+\i \frac{e}{\hbar}A_{\mu'}$, which brings in the four potential $A_{\mu'}\equiv (\Phi/c,-\vec{A})$, which contains the electric scalar potential $\Phi$ and the magnetic vector potential $\vec{A}$. 
In our considerations, we will treat these potentials as classical. 
With these alterations, the Dirac action (\ref{eqn:DiracAction1}) becomes
\begin{eqnarray}
S[\bar\psi',\psi']=\hspace{0.7\linewidth}\nonumber\\\int\bar\psi'(x^{\nu'})\left[\i \hbar \gamma^{\mu'}(u)\left(\p_{\mu'}+\i \frac{e}{\hbar}A_{\mu'}\right)-m\c\right]\psi'(x^{\nu'}) \qquad \label{eqn:DiracAction2}& &\\
\times \left(1+\frac{\g u}{\c^2}\right)\,\d {x'}^4\quad, \nonumber & &
\end{eqnarray}
which now is the action for a Dirac electron, in the presence of an electromagnetic field and seen by a homogeneously accelerated observer.   
However, an important feature of the system is still missing in Eq.~(\ref{eqn:DiracAction2}): the anomalous contribution to the magnetic moment of the electron. 
Therefore, we introduce the anomaly $a$, which accounts for the discrepancy between the gyromagnetic ratio ${\gs}_{\mathrm{Dirac}}=2$ of Dirac theory and the measured value $\gs=2(1+a)$, caused by interactions between the electron and the quantum vacuum.
In order to account for this anomaly, we introduce a non-minimal coupling of the electron to the electromagnetic field strength tensor $F_{\mu'\nu'}=\p_{\mu'}A_{\nu'}-\p_{\nu'}A_{\mu'}$.  
With this additional term the action reads 
\begin{eqnarray}
S[\bar\psi',\psi']=\hspace{0.7\linewidth}\nonumber\\\int\bar\psi'(x^{\nu'})\left[\i \hbar \gamma^{\mu'}(u)\left(\p_{\mu'}+\i \frac{e}{\hbar}A_{\mu'}\right)\right.\hspace{0.25\linewidth}& &\label{eqn:FinalAction}\\
\left.+\,\,\,a\,\frac{e\hbar}{2m\c}\frac{\i}{4}[\gamma^{\mu'}(u),\gamma^{\nu'}(u)]F_{\mu'\nu'}-m\c\right]\psi'(x^{\nu'}) \quad & & \nonumber\\
\times \left(1+\frac{\g u}{\c^2}\right)\,\d {x'}^4\quad, \nonumber & &
\end{eqnarray}
where the commutators of the coordinate dependent Dirac matrices are the generators of local Lorentz transformations. 
The structure of the additional term in the action can be motivated by QED considerations \cite{Fold58}. 
The action (\ref{eqn:FinalAction}) is discussed in detail for an electron in an inertial system in \cite{Brow86,Grae69,Bjor64} and references therein. Naturally, we recover their cases in the limit of vanishing acceleration $g$. 
%

As we discussed above, these steps were performed in order to derive Dirac equation in Rindler spacetime. 
Since we want to give this equation in the Hamiltonian representation, we separate the spacetime coordinates $(x^{\nu'})=(\c t,  x', y', u)$ into the time parameter $t$ and the spatial coordinates $(x^{i'})=(x', y', u)$, where $i'=1',\dots,3'$. 
Furthermore we redefine the Dirac matrices to be
\begin{equation}
\gamma^{0'}(u)=\left(1+\frac{\g u}{\c^2}\right)^{-1}\,\beta \quad,\quad \gamma^{i'}=\beta\alpha^{i'}\quad, \label{eqn:BetaAlpha1}
\end{equation}
where the \emph{constant} matrices  $\beta$ and $\alpha^{i'}$ obey the following relations:
\begin{equation}
\{\alpha^{i'},\alpha^{j'}\}=2\delta^{i'j'}\quad,\quad \beta\,\alpha^{i'}+\alpha^{i'}\beta=0\quad. \label{eqn:BetaAlpha2}
\end{equation} 
With these definitions (\ref{eqn:BetaAlpha1}) and (\ref{eqn:BetaAlpha2}) the relation (\ref{eqn:Clifford2}) for the $u$-dependent Dirac matrices is satisfied.
%

We combine the  $\alpha^{i'}$  to be a vector $\vec{\alpha}\equiv(\alpha^{x'},\alpha^{y'},\alpha^{u})$ and define the canonical momentum $\vec{\pi}\equiv\vec{p}-e\vec{A}$, where $\vec{p}\equiv -\i \hbar \nabla$ is the momentum operator and  $\nabla=(\p_{x'},\p_{y'},\p_u)$ is the gradient in the $(x',y',u)$-coordinate system. After these steps we rewrite the action
\begin{equation}
S[\psi'^\dagger,\psi']=\int \Bigl(\langle \psi' |\i \hbar \p_t \psi'\rangle-\langle \psi' | H\psi'\rangle \Bigr)\, \d t\quad. \label{eqn:FinalAction2}
\end{equation}
By variation of this action we directly obtain the Dirac equation in Schr\"odinger form 
\begin{equation}
\i \hbar \p_t |\psi'\rangle=  H|\psi'\rangle \quad.
\end{equation}
Moreover, the structure of Eq.~(\ref{eqn:FinalAction2}) helps us to construct the scalar product
\begin{equation}
\langle\psi_1'|\psi_2'\rangle\equiv \int \psi_1'^\dagger \psi_2' \left(1+\frac{\g u}{\c^2}\right)^{-1}\d x' \d y' \d u \quad. \label{eqn:ScalarProduct}
\end{equation}
and the corresponding hermitian Hamiltonian.
\begin{eqnarray}
H&\equiv&\left(1+\frac{\g u}{\c^2}\right)\left(\c\vec{\alpha}\cdot\vec{\pi}+m\c^2\beta-\,\frac{ae}{m}\beta\,\vec{B}\cdot\vec{s}\right)\label{eqn:Hamiltonian1}\\
& &\quad +\,e\Phi \,+\,\frac{\i ae\hbar}{2m\c}\beta \,\vec{\alpha}\cdot\vec{E}\quad\nonumber \,.
\end{eqnarray}
Here the electric and magnetic fields are defined by  $\vec{E}\equiv -\nabla \Phi-\p_t \vec{A}$ and $\vec{B}\equiv \nabla\times\vec{A}$. Moreover, $\vec{s}\equiv -\frac{\i}{4}\hbar\,\vec{\alpha}\times \vec{\alpha}$ is the four-spin operator.
It is easy to check, that the Hamiltonian (\ref{eqn:Hamiltonian1}) is hermitian with respect to the scalar product (\ref{eqn:ScalarProduct}).
%

\subsection{Non-relativistic reduction of the Hamiltonian}  \label{section:IIc}
In the last section, we derived the Hamiltonian of a relativistic spin 1/2 particle, that moves in Rindler spacetime in the presence of an electromagnetic field.
In addition we introduced the anomalous magnetic moment of this particle, which accounts for the interaction of the electron with the quantum vacuum.
The Hamiltonian (\ref{eqn:Hamiltonian1}), in its general form, can be applied for any velocities ($v<\c$) of an electron. 
In this work, however, we concentrate on a scenario, where the electron is stored in a Penning trap in a laboratory on Earth.
In this case, the velocity $v\ll \c$ is non-relativistic and the quantity $g L /c^2 \ll 1$  is a small parameter, where $L$ is a typical length scale of the experiment.
Therefore, we can simplify (\ref{eqn:Hamiltonian1}) to be the Hamiltonian of a non-relativistic particle and take into account correctional $1/\c^2$ effects only.
All higher orders in $1/\c$ are collected in the Landau symbol $\mathrm{O}(1/\c^3)$ and will be neglected later.
%

In what follows, we derive the non-relativistic reduction of the Hamiltonian (\ref{eqn:Hamiltonian1}).
There are many approaches to construct a non-relativistic Hamiltonian and its post-Newtonian corrections, see for example \cite{Giulini} and references therein.
In this work, we derive these corrections by a Foldy-Wouthuysen transformation \cite{Bjor64,Fold50}, which is used to decouple the electronic and positronic sector of $H$. 
The starting point of a Foldy-Wouthuysen transformation is to rewrite the Hamiltonian in the form  
\begin{equation}
H	= m\c^2 \beta + \mathcal{E} + \mathcal{O} \quad ,	\label{eqn:Hevenandodd}
\end{equation}
where the part of the Hamiltonian, which acts on the electronic and positronic degrees of freedom separately, is called the \emph{even} part of $H$:
\begin{eqnarray}
\mathcal{E}&=& m\g u\,\beta -\left(1+\frac{\g u}{\c^2}\right)\frac{ae}{m}\beta\,\vec{B}\cdot\vec{s}+\,e\Phi
\end{eqnarray}
and the part of the Hamiltonian, which couples the electronic and positronic sector is denoted as its \emph{odd} part:  
\begin{eqnarray}
\mathcal{O}&=& \left(1+\frac{\g u}{\c^2}\right)\c\vec{\alpha}\cdot\vec{\pi}+\,\frac{\i ae\hbar}{2m\c}\beta \,\vec{\alpha}\cdot\vec{E}\quad.
\end{eqnarray}
In the next step, we minimize the contribution of the odd part by an unitary transformation
\begin{eqnarray}
H' &=& \e^{\mathcal{W}}H\e^{-\mathcal{W}}\nonumber \\
   &=& H +[\mathcal{W},H]+\frac{1}{2!}[\mathcal{W},[\mathcal{W},H]]+\dots \quad ,	\label{eqn:BCH}
\end{eqnarray}
where the anti-hermitian operator $\mathcal{W}=\beta\, \mathcal{O}/(2m\c^2)$ is chosen, such that $[\mathcal{W},\beta\,m\c^2]=-\mathcal{O}$.
Because of this choice, the odd part $\mathcal{O}$ is canceled out in the first two terms in the right hand side of Eq. (\ref{eqn:BCH}). 
Considering all terms in (\ref{eqn:BCH}), which enter $H'$ up to order $\mathrm{O}(1/c^3)$, the new Hamiltonian can be written as
\begin{equation}
H'	=  m\c^2\beta + \mathcal{E}' + \mathcal{O}' \quad ,	
\end{equation}
in similarity to Eq. (\ref{eqn:Hevenandodd}). The new even and odd parts read
\begin{eqnarray}
\mathcal{E}'&=& \mathcal{E}+\frac{1}{2}[\mathcal{W},\mathcal{O}]+\frac{1}{2}[\mathcal{W},[\mathcal{W},\mathcal{E}]]\label{eqn:evenprime}\\
& &\qquad+\frac{1}{8}[\mathcal{W},[\mathcal{W},[\mathcal{W},\mathcal{E}]]]+\mathrm{O}(1/\c^3)\quad,\nonumber\\
\mathcal{O}'&=& [\mathcal{W},\mathcal{E}]+\frac{1}{3}[\mathcal{W},[\mathcal{W},\mathcal{O}]]+\mathrm{O}(1/\c^3)\quad.
\end{eqnarray}
We see, that the former odd part $\mathcal{O}$ appears in the commutators with $\mathcal{W}$ only, while the new odd part $\mathcal{O}'$ is proportional to $1/\c$ in the leading order.
To further minimize the order of $\mathcal{O}'$, we can iterate the Foldy-Wouthuysen transformation until  $\mathcal{O}'''=\mathrm{O}(1/\c^3)$  is reached, such that
\begin{eqnarray}
H'''&=&m\c^2 \beta +\mathcal{E}'''+\mathrm{O}(1/\c^3) \quad,
\end{eqnarray}
where the further iterations do not affect $\mathcal{E}'''=\mathcal{E}'+\mathrm{O}(1/c^3)$.
Therefore, by calculating (\ref{eqn:evenprime}), we find the Hamiltonian $H'''$ of the electron with anomalous magnetic moment and its antiparticle in an accelerated frame and an arbitrary electromagnetic field with all its corrections in $1/\c^2$:
\begin{eqnarray}
H'''&=&m \c^2\beta + um \g \beta + e\Phi \nonumber \hspace{10.63em}\textnormal{\emph{(i)}} \\
& & -\,\frac{1}{2m^2\c^2}\,\vec{s}\cdot \Bigl((1+2a)e\vec{E}-m\vec{\g}\beta\Bigr)\times\vec{\pi}\nonumber\hspace{2.61em}\textnormal{\emph{(ii)}}\\
& &  + \beta\left(1+\frac{\g u}{\c^2}\right)\left(\frac{1}{2m} \vec{\pi}^{\,2}-(1+a)\frac{e}{m}\,\vec{B}\cdot\vec{s}\right)\nonumber\quad\textnormal{\emph{(iii)}}\\
& &  -\,\frac{1}{8\c^2 m^3} \beta(\vec{\pi}^2-2e \vec{B}\cdot\vec{s})^2 \nonumber\\
& & +\,(1+2a)\frac{e\hbar^2}{8c^2m^2}\,\Delta\Phi\label{eqn:newHamilton}\\
& & +\,\frac{ae}{8m^3c^2}\beta\Bigl(2\{\vec{s}\cdot\vec{\pi},\vec{B}\cdot\vec{\pi}\} 
+[\vec{\pi}^2,\vec{B}\cdot\vec{s}]\Bigr)\nonumber\\
& &  +\,\frac{ae\hbar^2}{8 m^3 c^2}\,\beta\, (\nabla\times\vec{B})\cdot\vec{\pi} \nonumber\\
& &  +\,\frac{\i ae \hbar}{4 m^3 c^2} \beta\, [\nabla(\vec{B}\cdot\vec{s})]\cdot\vec{\pi}+\,\,\mathrm{O}(1/\c^3)\nonumber\quad,
\end{eqnarray}
where we defined the acceleration vector  $\vec{g}=(0,0,\g)$.
In the absence of gravity, i.e. for $\vec{g}=0$, all terms in $H'''$ are well known and have been studied extensively, first and foremost \cite{Brow86,Grae69} and references therein. The presence of gravity, however, gives rise to additional parts, which have to be discussed in more detail. As seen from Eq.~(\ref{eqn:newHamilton}), gravity enters this Hamiltonian at three points.
\emph{(i)} It gives the usual Newtonian potential, as it is known from classical mechanics,
\emph{(ii)} it  acts as a correction to the spin orbit coupling therm and 
\emph{(iii)} it induces a redshift of the non-relativistic kinetic energy and the coupling term between $\vec{B}$ and $\vec{s}$, which is important for our investigation of gravitational effects on free electron $\gs$-factor measurements. 
%

The Hamiltonian (\ref{eqn:newHamilton}) acts separately on the electronic and positronic degrees of freedom. Therefore, these two sectors are decoupled up to the desired order $\mathrm{O}(1/\c^3)$.
The choice of sector is made by selecting the positive or negative eigenvalue of $\beta$.
In our case we restrict ourselves to the discussion of the electron only, whose dynamics is described by $H'''$ after the replacement $\beta=+1$.
%

The Hamiltonian $H'''$ now contains all effects on the electron caused by the homogeneous acceleration and the anomalous magnetic moment and all relativistic effects, up to the order of $1/\c^2$.
It provides two particular limits, which are well known, either in the theory of Fermions in non-geodesic motion \cite{Fis81,Hehl90,Exp1,Exp2}, or in the physics of traps \cite{Brow86,Grae69}. 
In the case of vanishing electromagnetic fields, for example, we get the Hamiltonian of a free electron in accelerated motion:
\begin{eqnarray}
\lim \limits_{\vec{E},\vec{B}\to 0}
H'''&=&m \c^2 + um \g  
+ \left(1+\frac{\g u}{\c^2}\right)\frac{1}{2m} \vec{p}^{\,2} \label{eqn:LimitOne}\\
& &\quad -\,\frac{1}{8\c^2 m^3} \vec{p}^4  +\,\frac{1}{2m\c^2}\,\vec{s}\cdot ( \vec{\g}\times\vec{p})\nonumber\\
& &\hspace{11em}+\,\,\mathrm{O}(1/\c^3)\nonumber\quad.
\end{eqnarray} 
Here, the relativistic correction of the kinetic energy  $\sim \vec{p}^4$, the gravitational redshift  $(1+\g u/\c^2)$ and the  spin-gravity coupling $\sim \vec{s}\cdot(\vec{g}\times\vec{p})$ show up clearly. A detailed discussion of  (\ref{eqn:LimitOne})  and the physical consequences of the single terms can be found in \cite{Fis81,Hehl90,Exp1,Exp2}. On the other hand, in the absence of gravity the Hamiltonian (\ref{eqn:newHamilton})  reduces to
\begin{eqnarray}
\lim \limits_{\vec{g}\to 0}
H'''&=&m \c^2 + e\Phi \nonumber + \left(\frac{1}{2m} \vec{\pi}^{\,2}-(1+a)\frac{e}{m}\,\vec{B}\cdot\vec{s}\right)\nonumber\\
& & \quad -\,\frac{1}{8\c^2 m^3} (\vec{\pi}^2-2e \vec{B}\cdot\vec{s})^2\ \label{eqn:limit2}\\
& &\quad -\,(1+2a)\frac{e}{2m^2\c^2}\,\vec{s}\cdot( \vec{E}\times\vec{\pi})\nonumber\\
& &\quad +\,\frac{ae}{2 m^3c^2}(\vec{s}\cdot\vec{\pi})(\vec{B}\cdot\vec{\pi}) \,+\,\mathrm{O}(1/\c^3)\nonumber\quad,
\end{eqnarray}
where we, moreover, assumed a constant magnetic and a source free electric field.  Expression (\ref{eqn:limit2}) recovers \cite{Grae69} and has been the starting point for the investigation of relativistic corrections in geonium by Brown and Gabrielse \cite{Brow86} and Dehmelt \cite{Dehm88}, whose work we aim to extend by the consideration of a gravitational  field.
%

\section{The electron in a Penning Trap} \label{section:III}
\subsection{The electromagnetic field of a Penning trap\\ in Rindler spacetime} \label{section:IIIa}
In the previous section, we derived the Hamiltonian (\ref{eqn:newHamilton}) of an electron with anomalous magnetic moment, affected by a homogeneous gravitational and an arbitrary electromagnetic field. 
Although, this Hamiltonian can be applied to any kind of electric and magnetic fields, we want to apply it to the particular case of the electromagnetic field of a Penning trap.
Therefore we assume an ideal trap potential, without any imperfections, consisting of a static homogeneous magnetic field and a static electric quadrupole field. 
Moreover, the Penning trap is placed in the spacetime of an accelerated observer. Therefore, the electromagnetic field of the trap is distorted.
In order to account for this distortion, we need to formulate Maxwell equations in Rindler spacetime. 
These equations for a source free electromagnetic field in their covariant form read   
\begin{equation}
\p_{\mu'}F^{\mu'\nu'}+\Gamma^{\mu'}_{\mu'\rho'}F^{\rho'\nu'}=0 \quad, \label{eqn:Maxwell}
\end{equation}
in terms of the electromagnetic field strength tensor $F_{\mu'\nu'}=\p_{\mu'}A_{\nu'}-\p_{\nu'}A_{\mu'}$ and the Christoffel symbols $\Gamma^{\mu'}_{\mu'\rho'}=\frac{1}{2}g^{\mu'\sigma'}\p_{\rho'}g_{\mu'\sigma'}=\frac{\g}{\c^2}\left(1+\frac{\g u}{\c^2}\right)^{-1}\, \delta^{3}_{\rho'}$. For more details and the definition of the Christoffel symbols, see Appendix B.
%

Eq.~(\ref{eqn:Maxwell}) allows us to calculate the vector potential $\vec{A}$ and the scalar potential $\Phi$, needed in the Hamiltonian (\ref{eqn:newHamilton}). These potentials enter into the four-potential $(A_{\mu'})=(\Phi/\c,-\vec{A})$, which we assume to be static, i.e. $\p_t A_{\mu'}=0$ and, moreover, to satisfy the gauge condition 
\begin{equation}
\p_{\mu'}A^{\mu'}+\Gamma^{\mu'}_{\mu'\rho'}A^{\rho'}=0 \quad. \label{eqn:gauge}
\end{equation}
Under these assumptions, we can rewrite the spatial part of equation  (\ref{eqn:Maxwell}) in the form
\begin{eqnarray}
\nabla\cdot\left[\left(1+\frac{\g u}{\c^2}\right)\nabla \vec{A}\right]&=&\frac{\g}{c^4}\left(1+\frac{\g u}{\c^2}\right)^{-1} \vec{g}\,A_{3} \label{eqn:MaxwellA}\quad,
\end{eqnarray}
which determines the vector potential $\vec{A}=(A_1,A_2,A_3)$.
In this expression, moreover, $\nabla=(\p_{x'},\p_{y'},\p_u)$ is the gradient in the coordinate system $(x',y',u)$.
In order to solve Eq. (\ref{eqn:MaxwellA}), one has to define explicit boundary conditions. In the case of a Penning trap configuration, for example, we demand that $\vec{A}$ is the vector potential of a constant magnetic field $\vec{B}^{(0)}=(B_1,B_2,B_3)$ in the center of the trap. For this requirement, the solution of (\ref{eqn:MaxwellA}) is given by
\begin{eqnarray}
\vec{A}&=&-\frac{1}{2}\left(\begin{array}{c}
x' \\
y' \\
w(u)
\end{array}\right)\times \left(\begin{array}{c}
B_1\left(1+\frac{\g u}{\c^2}\right)^{-1}\\
B_2\left(1+\frac{\g u}{\c^2}\right)^{-1}\\
B_3\\
\end{array}\right)\,, \label{eqn:RindlerA}
\end{eqnarray}
where
\begin{eqnarray}
w(u)& =& \frac{\c^2}{\g}\left(1+\frac{\g u}{\c^2}\right)\,\log \left(1+\frac{\g u}{\c^2}\right)\,. 
\end{eqnarray}
For vanishing acceleration $\vec{g}=0$, Eq.~(\ref{eqn:RindlerA}) reduces to the well known vector potential of a constant magnetic field $-\frac{1}{2} \, \vec{r}'\times \vec{B}^{(0)}$ globally.
However, the presence of gravity leads to a distortion of the magnetic field, which is characterized by the factor $(1+g u /\c^2)$.
%

In the same way, we can find the scalar quadrupole potential $\Phi$ of the trap. This potential has to be a solution to the time-component of Eq. (\ref{eqn:MaxwellA}), which under gauge condition (\ref{eqn:gauge}) becomes
\begin{equation}
\nabla\cdot\left[\left(1+\frac{\g u}{\c^2}\right)^{-1}\!\!\nabla\Phi\right]=0\quad. \label{eqn:MaxwellPhi}
\end{equation}
Again the physically relevant boundary conditions have to be set here.
They are chosen such that the potential $\Phi$ is determined by a constant, traceless quadrupole matrix $\hat Q$ in the coordinate center. The corresponding solution to Eq. (\ref{eqn:MaxwellPhi}) is given by
\begin{eqnarray}
\Phi&=& \Bigl(x',y',f(u)\Bigr)\cdot \hat{Q}\cdot  \left(\begin{array}{c} 
x'\\
y'\\
f(u)
\end{array}\right)+ Q_{33} \,h(u)\,,\label{eqn:RindlerPhi}
\end{eqnarray}
where
\begin{eqnarray}
f(u)&=&\left(1+\frac{\g u}{2 \c^2}\right)\, u\,
\end{eqnarray}
and
\begin{eqnarray}
h(u)&=&\left(\frac{\c^2}{\g}\right)^2\left\{\left(1+\frac{\g u}{\c^2}\right)^2\log\left(1+\frac{\g u}{\c^2}\right)\right.\nonumber\\
& &  \left.- \frac{\g u}{\c^2}\left(1+\frac{\g u}{2 \c^2}\right)\left[1+\frac{\g u}{\c^2}\left(1+\frac{\g u}{2 \c^2}\right)\right]\right\}\,.
\end{eqnarray}
For vanishing acceleration $\vec{g}=0$, the potential (\ref{eqn:RindlerPhi}) reduces to the ideal quadrupole potential $\vec{r}'\cdot (\hat Q \cdot \vec{r}')$.
%

Having found the solutions  (\ref{eqn:RindlerA}) and  (\ref{eqn:RindlerPhi}) for the vector and scalar potential, we are ready now to set up the Penning trap field configuration.
For this purpose, we need to specify the geometry of the trap.
We adopt the coordinate system such that $B_2=0$ and introduce the angle $\theta$ between $\vec{B}^{(0)}$ and $\vec{g}$. 
For this choice, the constants $\vec{B}^{(0)}$ and $\hat Q$ are given by
\begin{equation}
\vec{B}^{(0)}=(-B \,\sin \theta\,,\,0\,,\,B\,\cos\theta)\quad
\end{equation}
and
\begin{equation}
\hat Q = -\frac{V}{4L^2}\left(\begin{array}{ccc}
\frac{1}{2}(3\cos(2\theta)-1)&0& 3 \cos\theta\sin\theta\\
0& 1 & 0 \\
3 \cos\theta\sin\theta & 0 & -\frac{1}{2}(3\cos(2\theta)+1)
\end{array}\right)\quad ,
\end{equation}
where $B$ is the absolute value of $\vec{B}^{(0)}$ and the quantities $V$ and $L$ are the typical voltage and spatial length scale of the Penning trap.
Having found the vector and scalar potentials $\vec{A}$ and $\Phi$, we can use them in the exact Dirac Hamiltonian (\ref{eqn:Hamiltonian1}), or in its expanded form (\ref{eqn:newHamilton}), as we will do in the next section.
%

\subsection{The electron in a Penning trap in Newtonian gravity} \label{section:IIIb}
We derived the Hamiltonian $H'''$  of an electron with anomalous magnetic moment in an accelerated frame and an arbitrary electromagnetic field in Sec. \ref{section:IIc}. Below, we want to use this Hamiltonian to describe the electron dynamics in a Penning trap, distorted by acceleration. Therefore, we insert the vector and scalar potentials (\ref{eqn:RindlerA}) and (\ref{eqn:RindlerPhi}) into Eq. (\ref{eqn:newHamilton}). For the sake of brevity, we will not present this lengthy expression here, that contains all relativistic effects on both, the electron and the trap, up to the order of $1/\c^2$. 
In this section, we derive the exact solution of the eigenvalue problem of the Hamiltonian in the Newtonian limit of low velocities and weak gravitational fields. Treating $1/\c^2$-effects as first order perturbations, we find the solution of the whole eigenvalue problem of $H'''$, afterwards.
Within the Newtonian limit, the non-relativistic Hamiltonian $H_0$ is obtained by considering the zeroth order in an  $1/\c$-expansion of $H'''$, only:
\begin{equation}
H'''=H_0+\,\mathrm{O}(1/\c)\quad,
\end{equation}
where
\begin{eqnarray}
H_0&=&m\c^2+ m\,\vec{\g}\cdot\vec{r}'+e\,\vec{r}'\cdot(\hat{Q}\cdot\vec{r}')\label{eqn:H0}\\
& & \qquad\qquad +\frac{1}{2m} \vec{\pi}^2-\frac{e\gs }{2m} \vec{B}^{(0)}\cdot\vec{s}\quad. \nonumber
\end{eqnarray}
In this expression we dropped the auxiliary coordinate $u$ in favour of the coordinate system $\vec{r}'=(x',y',z')$, given by Eq. (\ref{eqn:firstRindlerMetric}). Since $u=z'+\mathrm{O}(1/\c)$, this coordinate transformation allowed us to replace the canonical momentum by $\vec{\pi} = \vec{p}' + \frac{1}{2}\,e \, \vec{r}'\times \vec{B}^{(0)} $ and the electromagnetic potentials by $\Phi=\vec{r}'\cdot (\hat Q \cdot \vec{r}')$ and $\vec{A} = - \frac{1}{2} \, \vec{r}'\times \vec{B}^{(0)} $.
%

%
\begin{figure}[b!]
	\centering
	\includegraphics[width=0.5\linewidth]{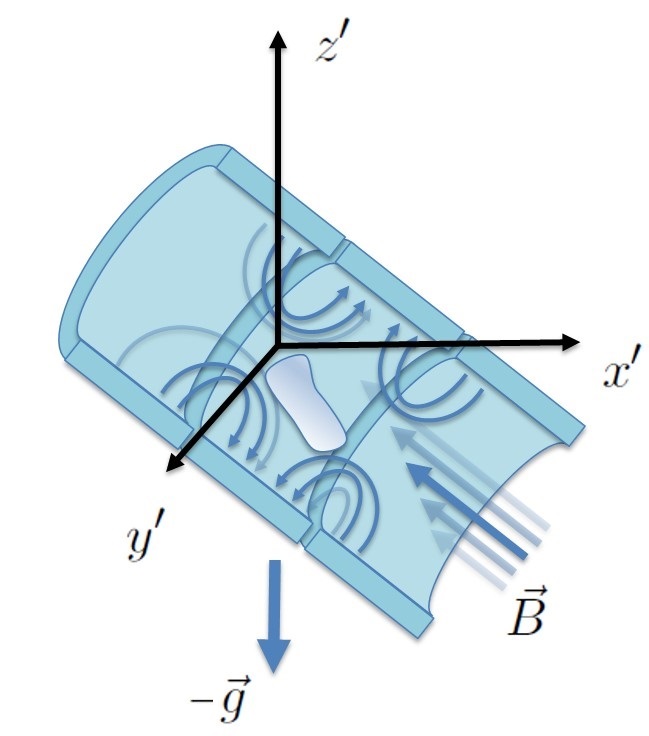}\includegraphics[width=0.5\linewidth]{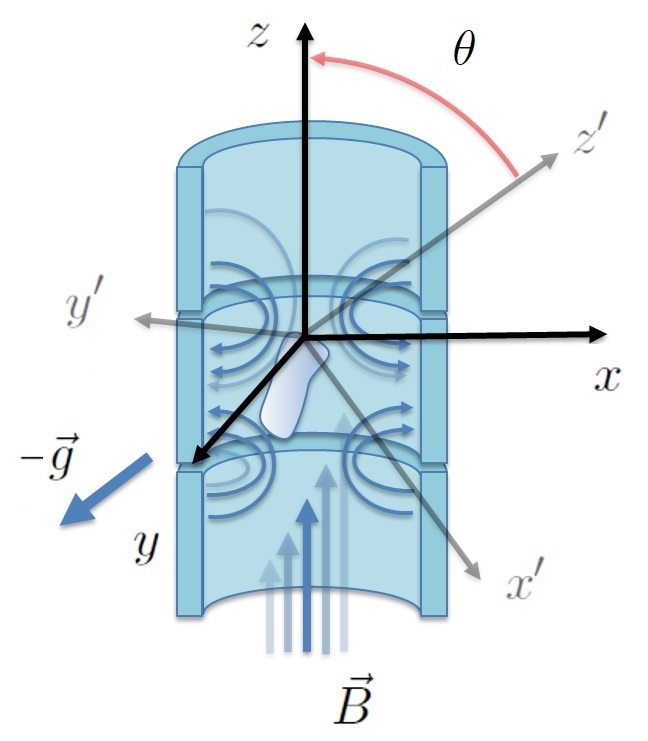}
	\caption{(Color online) from left to right: Change from laboratory frame with coordinates $(x',y',z')$, where $\vec{g}$ points into $z'$-direction, to the frame of trap geometry with $(x,y,z)$, where the $z$-axis and $\vec{B}^{(0)}$ are aligned. The angle $\varbeta$ is determined by the relation $\vec{g}\cdot\vec{B}^{(0)}=\g B \cos\varbeta$. }
	\label{Fig1}
\end{figure}
\begin{figure}[t!]
	\centering
	\includegraphics[width=0.5\linewidth]{Bild4.jpg}\includegraphics[width=0.5\linewidth]{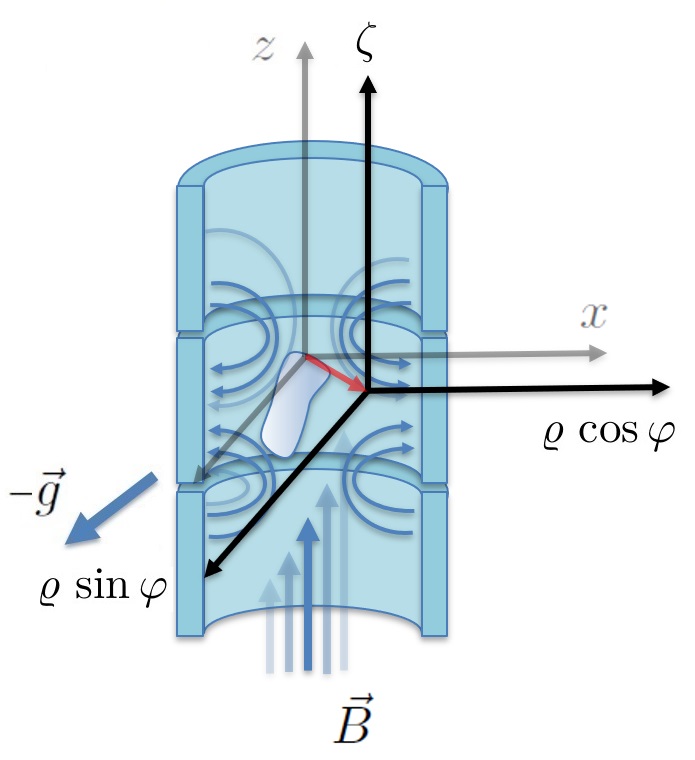}
	\caption{(Color online) After the rotation, shown in FIG.~\ref{Fig1}, the coordinate center is shifted into the new equilibrium position of the electron motion. }
	\label{Fig2}
\end{figure}

Eq. (\ref{eqn:H0}) closely resembles the well known Hamiltonian of a non-relativistic electron in a Penning trap \cite{Brow86}. The essential difference is the presence of the Newtonian potential $m\,\vec{\g}\cdot\vec{r}'$. The Hamiltonian $H_0$ can be further simplified by performing two additional transformations. First, we rotate the coordinate system, such that the $z$-axis is aligned with the direction of the magnetic field $\vec{B}^{(0)}$, see FIG.~\ref{Fig1}.  This is conventional in the analysis of Penning trap experiments \cite{Brow86,Dehm88}. As a second step, we shift the coordinate center by a constant vector, such that it coincides with the new equilibrium position of the electron motion, see FIG.~\ref{Fig2}. Moreover, we apply an unitary transformation $\tilde H_0= U^\dagger H_0 U$ in order to shift the momentum in $y$-direction by a constant value.
While a detailed discussion of these transformations is given in Appendix \ref{section:C}, here we just present the obtained Hamiltonian
\begin{eqnarray}
\tilde H_0&=& m\c^2+ \frac{m}{2}\omega_z^2\,\zeta^2 +\frac{m}{8}(\omega^2_c-2\omega_{z}^2)\,\rho^2 +\frac{1}{2m} \vec{P}^2 \qquad \label{eqn:hatH0}\\
& & \quad-\frac{\omega_c}{2}(L_3+\gs\, s_3)+\frac{m\g^2}{\omega_z^2}\left(1-\frac{3}{2}\cos^2\theta\right) \,.\nonumber 
\end{eqnarray}
Here, due to the axial symmetry of a Penning trap, it is convenient to use cylindrical coordinates $\vec{R} = 
( \rho, \varphi, \zeta)$. The momentum operator in this coordinate system is denoted as $\vec{P}$. In Eq.~(\ref{eqn:hatH0}), moreover, 
\begin{eqnarray}
\omega_c & =&  eB/m \quad, \label{eqn:cyclotron}\\
\omega_z & = &  \sqrt{eV/(mL^2)}\label{eqn:axial}\quad.
\end{eqnarray}
are the cyclotron frequency $\omega_c$ and the axial frequency  $\omega_z$.
%

The operator $\tilde H_0$ now is the well known Hamiltonian of an electron in a Penning trap \cite{Brow86}, except for the very last term in Eq. (\ref{eqn:hatH0}). This term describes the effect of Newtonian gravity on the electron and depends on the orientation of the Penning trap with respect to the acceleration $\vec{g}$. Since this term is constant, we are able to solve the eigenvalue problem 
\begin{equation}
\tilde H_0 \phi_{0}^{k,n,\ell,s} = E_{k,n,\ell,s} \phi_{0}^{k,n,\ell,s} \label{eqn:eigenvalueeq}
\end{equation}
analytically. As the result, we get the well established energies of the eigenvalue problem of geonium \cite{Brow86}, shifted by this constant gravitational term:
\begin{eqnarray}
& & E_{k,n,\ell,s}=m\c^2\nonumber\\
& &\quad +\,\hbar \omega_z \left(k+\frac{1}{2}\right)+ \hbar \omega_{c'} \left(n+\frac{1}{2}\right)-\hbar \omega_{m}\left(\ell+\frac{1}{2}\right)\nonumber\\
& &\quad +
\,\frac{\gs}{2}\,\hbar\omega_{c}\,s+\frac{m\g^2}{\omega_z^2}\left(1-\frac{3}{2}\cos^2\theta\right)\,. \label{eqn:Energy}
\end{eqnarray}
Here, following \cite{Brow86}, $k$ and $n$ are the non-negative integer quantum numbers of axial and cyclotron oscillation, while  $\ell=0,1,2,\dots$ and $s=\pm 1/2$ account for the angular momentum and spin projection of the electron onto the direction of the magnetic field. Moreover, the corrected or reduced cyclotron frequency $\omega_{c'}$ and the magnetron frequency $\omega_m$ are defined by
\begin{eqnarray}
\omega_{c'} &=& (\omega_{c} + \sqrt{\omega^2_c-2\omega_{z}^2} \,)/2  \quad, \label{eqn:cprime}\\
\omega_m &=&(\omega_{c} - \sqrt{\omega^2_c-2\omega_{z}^2} \,)/2\quad .  \label{eqn:magnetron}
\end{eqnarray}
Together with the axial frequency $\omega_z$, these are standard observables in Penning trap experiments.
%

As seen from (\ref{eqn:Energy}), Newtonian gravity leads to a constant shift of energy levels, only. This shift is independent of the quantum numbers of the electron in the trap and, therefore, does not effect frequencies of bound-bound transitions in geonium. In the next section we will see, that this is not the case if we take into account relativistic effects.  
%

\subsection{Relativistic energy correction for a gravitationally influenced electron in a Penning trap} \label{section:IIIc}
In the previous section we considered the Hamiltonian of a non-relativistic electron in a Penning trap in the presence of a homogeneous Newtonian gravitational field. The eigenvalues of this Hamiltonian are given by (\ref{eqn:Energy}), while the explicit form of corresponding eigenfunctions is given in Appendix \ref{section:C}. In this section we will use these eigenfunctions as a basis for a perturbation analysis in order to account for relativistic effects. The perturbation $\tilde H_I$ of the Hamiltonian $\tilde H_0$ can be formally written as
\begin{equation}
\tilde H_I=\tilde H'''-\tilde H_0+\,\mathrm{O}(1/\c^3)\quad.
\end{equation}
Within first order perturbation theory, the energy corrections can be expressed by
\begin{equation}
\delta E_{k,n,\ell,s}=\langle \phi^{k,n,\ell,s} |(\tilde H'''-\tilde H_0)|\phi^{k,n,\ell,s}\rangle +\,\mathrm{O}(1/\c^3)\,, \label{eqn:corr}
\end{equation}
where similar to the steps, leading to Eq. (\ref{eqn:hatH0}), we apply a transformation to cylindrical coordinates $\vec{R} = ( \rho, \varphi, \zeta)$ and perform the unitary transformation  $\tilde H'''=\tilde H'''(\vec{R},\vec{P})=U^\dagger H'''(\vec{r}',\vec{p}')U$, afterwards. However, in contrast to the last section, the auxiliary coordinate $u=z'-\g z'^2/(2c^2)+\,\mathrm{O}(1/\c^3)$ is now replaced by $z'$, taking into account the relativistic corrections of order $1/\c^2$.
%

While the energy correction (\ref{eqn:corr}) can be applied for any set of quantum numbers $n$, $k$, $\ell$, $s$, in the following we want to use these energy corrections in order to investigate the relativistic effects on $\gs$-factor measurements, as they are discussed in \cite{Gab08}. In these experiments the transitions between the lowest energy levels in geonium are driven under the change of quantum numbers $n$ and $s$, while $k=\ell=0$. For this scenario the energy correction reads
\begin{eqnarray}
\delta E_{0,n,0,s}/\hbar	&=&	-\frac{1}{8}(1+2n+2s)^2\delta \label{eqn:energyshifts}\\
& & -\frac{1}{2}(1+2n+\gs s)\sigma_1(\theta)+(1+n)\sigma_2(\theta)\,,\quad\nonumber 
\end{eqnarray}
where $\delta$ is the special relativistic correction due to cyclotron motion:
\begin{eqnarray}
\delta			&\equiv &	\frac{\hbar\omega_c^2}{m c^2}\,. \label{eqn:delta}
\end{eqnarray}
Moreover, the frequencies 
\begin{eqnarray}
\sigma_1(\theta)	&\equiv &	-2\,\frac{\g^2\omega_c}{\c^2\omega_z^2}\,\cos^2\theta\left(1-\frac{3}{2}\cos^2\theta\right)\,\label{eqn:sigma1}
\end{eqnarray}
and
\begin{eqnarray}
\sigma_2(\theta)	&\equiv &	\frac{\g^2}{2\c^2\omega_c}\,\sin^2\theta\left(1-\frac{9}{4}\sin^2(2\theta)\right)\, \label{eqn:sigma2}
\end{eqnarray}
are related to the first order non-vanishing gravitational effects. In these expressions we assumed $\omega_c=\omega_{ c}'$ and neglected all higher orders of  $\omega_z/\omega_c$. In order to investigate the corrections to
the $\gs$-factor formula introduced by \cite{Brow86} we will use the energy correction (\ref{eqn:energyshifts}) in the next section.
%

\subsection{Gravitational effect on free electron $\gs$-factor measurements} \label{section:IIId}
\begin{figure}[t!]
	\centering
	\includegraphics[width=0.7\linewidth]{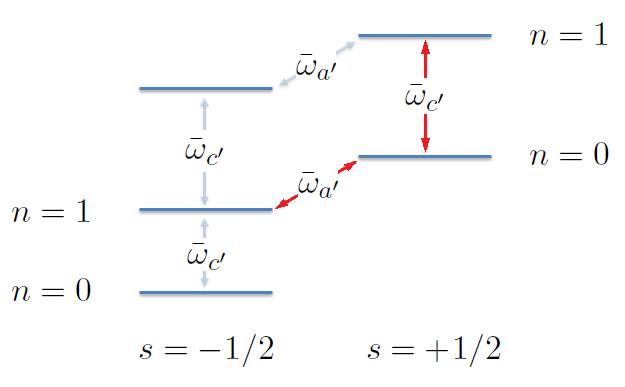}
	\caption{(Color online) Level scheme for spin states and the lowest cyclotron oscillator states of an electron in a Penning trap.
		Here, $n$ is the quantum number of cyclotron oscillation and $s=\pm 1/2$ refers to the two spin states of the electron in the Penning trap.
		The spin-flip transition $\bar{\omega}_{a'}$ (\ref{eqn:anomalous}) and the cyclotron transition $\bar{\omega}_{c'}$ (\ref{eqn:cyclo}), used to determine the free electron $\gs$-factor are marked in red. Due to the $n$,$s$-dependence of the relativistic corrections (\ref{eqn:energyshifts}), the energy levels are not equidistant.
	}
	\label{Fig4}
\end{figure}

Having derived the energy (\ref{eqn:Energy}) of an electron in a Penning trap and its relativistic correction (\ref{eqn:energyshifts}), we are prepared to discuss the effect of gravity on the result of free electron $\gs$-factor measurements. 
Therefore, we follow the steps, performed by L.~S.~Brown and G.~Gabrielse  in order to obtain the $\gs$-factor formula, presented in \cite{Brow86}, which in our case will contain additional corrections. 
In their analysis, the $\gs$-factor is extracted by the measurement of two frequencies of transitions in geonium, namely the anomalous frequency and the reduced cyclotron frequency.  
The first one we obtain from the spin-flip transition between the energy levels ($n=1$, $s=-1/2$) and ($n=0$, $s=+1/2$), see FIG.~\ref{Fig4}. By employing Eq.~(\ref{eqn:Energy}) and Eq.~(\ref{eqn:energyshifts}), this frequency can be calculated as
\begin{eqnarray}
\bar{\omega}_{a'} &=& (E+\delta E)_{0,0,0,+1/2}/\hbar - (E+\delta E)_{0,1,0,-1/2}/\hbar \nonumber \\
&=& \gs \omega_c/2 -\omega_{c'} - (\gs/2-1) \sigma_1 (\theta) - \sigma_2 (\theta) \quad.\label{eqn:anomalous}
\end{eqnarray}
In contrast to previous investigations \cite{Gab08,Gab06,Brow86}, this expression now also contains the gravitational correction frequencies $\sigma_1 (\theta)$ and $\sigma_2 (\theta)$. 
In order to extract the free electron $\gs$-factor from Eq.~(\ref{eqn:anomalous}), we can rearrange it
\begin{equation}
\gs/2=1+\frac{{\bar{\omega}}_{ a'}-(\omega_c-\omega_{c'}) + \sigma_2(\theta)}{\omega_c-\sigma_1(\theta)}\quad \label{eqn:anomaly01}
\end{equation}
and further simplify it, using
\begin{equation}
\omega_c-\omega_{ c'} =\omega_m = \frac{\omega_z^2}{2\omega_{c'}} \quad, \label{eqn:simp}
\end{equation}
which can be obtained from the definitions of the frequencies (\ref{eqn:cyclotron}),(\ref{eqn:axial}) and (\ref{eqn:cprime}), (\ref{eqn:magnetron}).
With the help of Eq.~(\ref{eqn:simp}), we obtain
\begin{equation}
\gs/2=1+\frac{{\bar{\omega}}_{ a'}-\frac{\omega_z^2}{2\omega_{c'}} + \sigma_2(\theta)}{\omega_{c'}+\frac{\omega_z^2}{2\omega_{c'}}-\sigma_1(\theta)} \label{eqn:anomaly02}\quad.
\end{equation}
As seen from this expression, the $\gs$-factor formula depends not only on ${\bar{\omega}}_{ a'}$, but also on $\omega_{c'}$ and $\omega_z$.
Since the latter is obtained by tracking the mirror charge of the electron \cite{Brow86}, we assume $\omega_z$ to be known and focus on the discussion of $\omega_{c'}$.
%

In the non-relativistic limit, $\omega_{c'}$ is the frequency of transitions between the energy levels ($n=0$, $s=1/2$) and ($n=1$, $s=1/2$).
In practice, the frequency of this transition is affected by relativistic effects, see FIG.~\ref{Fig4}. This actually measured frequency will be denoted as $\bar{\omega}_{c'}$. Using Eq.~(\ref{eqn:Energy}) and Eq.~(\ref{eqn:energyshifts}), we can express this frequency as
\begin{eqnarray}
\bar{\omega}_{c'} &=& (E+\delta E)_{0,1,0,1/2}/\hbar - (E+\delta E)_{0,0,0,1/2}/\hbar \nonumber \\
&=&{\omega}_{c'} - \frac{3}{2}\delta -\sigma_1 (\theta) + \sigma_2 (\theta) \quad.
\end{eqnarray}
By this relation, we can express ${\omega}_{c'}$ in Eq.~(\ref{eqn:anomaly02}) in terms of $\bar{\omega}_{c'}$ and the correction frequencies. In order to further simplify the $\gs$-factor formula, we make a Taylor expansion of 
\begin{eqnarray}
\frac{\omega_z^2}{2\omega_{c'}} &=& \frac{\omega_z^2}{2({\bar{\omega}}_{c'}+3\delta/2+\sigma_1 (\theta) - \sigma_2 (\theta)) }\label{eqn:cyclo}\\
&\approx& \frac{\omega_z^2}{2\bar{\omega}_c'}- \frac{1}{2}\left(\frac{\omega_z}{\omega_c}\right)^2\left(\frac{3}{2}\delta +\sigma_1 (\theta) -\sigma_2 (\theta)\right)  \quad. \nonumber
\end{eqnarray}
In the last step, we insert this expansion in Eq.~(\ref{eqn:anomaly02}), where we consider only the leading contributions of $\delta$, $\sigma_1(\theta)$ and $\sigma_2(\theta)$. We finally obtain 
\begin{eqnarray}
\gs/2	&=&	1+\frac{\bar{\omega}_{a'}+\sigma(\theta)-\omega_z^2/(2\bar{\omega}_{c'})}{\bar{\omega}_{c'}+\frac{3}{2}\delta-\sigma(\theta)+\omega_z^2/(2\bar{\omega}_{c'})}\quad, \label{eqn:final}
\end{eqnarray}
where all gravitational correction frequencies enter the expression
\begin{equation}
\sigma(\theta)= \frac{1}{2}\left(\frac{\omega_{z}}{\omega_{c}}\right)^2\sigma_1(\theta)+\sigma_2(\theta).
\end{equation}
In the case of vanishing acceleration, i.e. $\vec{g}=0$, this equation recovers the known  $\gs$-factor formula  by L.~S.~Brown and G.~Gabrielse from \cite{Brow86}, while the presence of gravity leads to additional contributions to this formula.
In order to illustrate this, we can expand Eq.~(\ref{eqn:final}) into
\begin{equation}
\gs/2=\gs^{(0)}/2+\delta_\sigma\gs/2\quad, 
\end{equation}
where $\gs^{(0)}$ is the known expression for the free electron $\gs$-factor from \cite{Brow86}, while the relative shift of $\gs$-factor due to gravity is of the order
\begin{equation}
\frac{\delta_\sigma \gs}{\gs}\sim \frac{1}{2} \frac{(\g/\c)^2}{\omega_c^2}\quad.
\end{equation}
In the recent Penning trap experiments of \cite{Gab08}, a cyclotron frequency of  $\omega_c=2\pi\cdot 149 \,\mathrm{GHz}$ is used.
For such an experiment performed in a laboratory on Earth, i.e. $\g=9.81\, \mathrm{m}/\mathrm{s}^2$, we obtain $\delta_\sigma \gs/\gs\sim 6.1\times 10^{-40} $.
%

\section{Summary and Conclusion} \label{section:IV}
In this work we presented a theoretical investigation of an electron in a Penning trap in the presence of a gravitational field.
In this system we analyzed how the presence of gravity may affect the result of free electron $\gs$-factor measurements.
Therefore, we considered a single electron with anomalous magnetic moment in the presence of electromagnetic fields in the spacetime of homogeneous acceleration. 
For this scenario we derived the Hamiltonian (\ref{eqn:newHamilton}), which accounts for the relativistic effects up to order $1/\c^2$.
This Hamiltonian has been applied to the electron dynamics in a gravitational distorted Penning trap, whose electromagnetic field (\ref{eqn:RindlerA}), (\ref{eqn:RindlerPhi}) is given as an exact solution of Maxwell equations in Rindler spacetime.
Making use of first order perturbation theory, we derived analytical expressions for the energy eigenvalues (\ref{eqn:Energy}),(\ref{eqn:energyshifts}) of that Hamiltonian up to order $1/\c^2$.
%

A detailed analysis of these energies has shown, that Newtonian gravity only leads to constant shifts of the energy levels of geonium. Thus, Newtonian gravity has no effect on measured transition frequencies.
In contrast, the relativistic effects of order  $1/\c^2$ lead to relative shifts of the energy levels.
We, therefore, argue that these relativistic corrections may affect the $\gs$-factor measurements, which rely on transitions in geonium.
In order to quantify the gravitational effects, we derived the expression (\ref{eqn:final}), which for $\vec{\g}=0$ recovers the known $\gs$-factor formula introduced by L.~S.~Brown and G.~Gabrielse, while for $\vec{\g}\not=0$ it predicts a shift of the measured $\gs$-factor of $\delta_\sigma \gs/\gs\sim 6.1\times 10^{-40}$.
While this can not be measured in experiments of current accuracy, it can be enhanced in the case of lower frequencies and higher accelerations and, therefore, may be important for future studies.
%

\section*{Appendix}

\appendix

\section{Coordinate transformation of Dirac action towards Rindler spacetime} \label{section:A}
In section \ref{section:IIb} we pointed out, that some attention has to be drawn to the transformation of the Dirac action towards Rindler spacetime.
The generalization of Dirac equation to curved spacetime or spacetime of non-geodesic motion is discussed in a wide range of publications, see for instance \cite{Hehl90,Lippoldt,Weldon}.
 Indeed there are many degrees of freedom -- especially the freedom of an additional spin base transformation of the spinor and/or the Dirac matrices. 
In this Appendix, we show the way we have chosen to get the Dirac action in the form of (\ref{eqn:DiracAction1}), where the spin base of the spinor and the Dirac matrices transforms under the spin representation of the coordinate transformation (\ref{eqn:RindlerTrafo}).
We want to emphasize, that this is a choice, that is of advantage in our case, and by no means an advice how to perform such a transformation in general.  
%

Starting with the Dirac action (\ref{eqn:DiracAction}), we perform the coordinate transformation (\ref{eqn:RindlerTrafo}) in the form of $x^{\mu}=\frac{\p x^{\mu}}{\p x^{\mu'}}\,x^{\mu'}$, where spacetime indices transform under the common properties of coordinate differentials. For the derivative and the volume of the spacetime integral, this means 
\begin{equation}
\p_{\mu}=\frac{\p x^{\mu'}}{\p x^{\mu}}\,\p_{\mu'}\quad,\quad \d x^4=\left(1+\frac{\g u}{\c^2}\right)\,\d {x'}^4\quad. \label{eqn:A01}
\end{equation}
In addition we want to allow, that the spinor and the Dirac matrices are affected by a spin base transformation $\S=\S(x^{\mu'})$, which has to be specified later:
\begin{equation}
\psi(x^{\mu})=\S\psi'(x^{\mu'})\quad,\quad\gamma^\mu=\frac{\p x^{\mu}}{\p x^{\mu'}} \S\gamma^{\mu'}(u)\S^{-1}\quad.
\end{equation} 
Under these assumptions, the Dirac adjoint spinor reads
\begin{eqnarray}
\bar\psi(x^{\mu})&=&(\psi(x^{\mu}))^\dagger \gamma^0 \label{eqn:A03}\\
&=&(\psi'(x^{\mu'}))^{\dagger}\S^{\dagger}\frac{\p x^{0}}{\p x^{\mu'}} \S\gamma^{\mu'}(u)\S^{-1} \,. \nonumber
\end{eqnarray} 
Applying the transformations (\ref{eqn:A01}) - (\ref{eqn:A03}) to the Dirac action, we obtain
\begin{eqnarray}
S[\bar\psi',\psi']=\int(\psi'(x^{\nu'}))^\dagger \S^\dagger \S \frac{\p x^{0}}{\p x^{\mu'}}\gamma^{\mu'}(u) \hspace{0.15 \linewidth} & & \label{eqn:A04}\\ \times \,\left[\i \hbar \gamma^{\mu'}(u)\left(\p_{\mu'}+S^{-1}\p_{\mu'}\S\right)-m\c\right]\psi'(x^{\nu'}) & &\nonumber\\
\times \,\, \left(1+\frac{\g u}{\c^2}\right)\,\d {x'}^4\quad, \nonumber & &
\end{eqnarray}
where we used $\frac{\p x^{\mu}}{\p x^{\mu'}}\frac{\p x^{\nu'}}{\p x^{\mu}}=\delta^{\nu'}_{\mu'}$ and $\S\S^{-1}=1$.
In similarity to $\bar\psi(x^{\mu})=(\psi(x^{\mu}))^\dagger \gamma^0 $, we want the Dirac adjoint spinor to have the form $\bar\psi'(x^{\nu'})=(\psi'(x^{\mu'}))^\dagger  \gamma^{0'}(u)$ in the new coordinate system. This leads to the condition
\begin{equation}
\S^\dagger \S \frac{\p x^{0}}{\p x^{\mu'}}\gamma^{\mu'}(u) \stackrel{!}{=}
\gamma^{0'}(u)\quad, \label{eqn:determineS}
\end{equation}
which is suitable to determine $\S^\dagger \S $, fixing $\S$ up to an unitary transformation.
The used spin base transformation, which satisfies (\ref{eqn:determineS}) is
\begin{eqnarray}
\S=& &\sqrt{1+\frac{\g u}{\c^2}}\,\cosh\left(\frac{\g t}{2c}\right)\, \gamma^{0'}(u)\\
& &\qquad+ \,\frac{1}{\sqrt{1+\frac{\g u}{\c^2}}}\,\sinh\left(\frac{\g t}{2c}\right)\, \gamma^{3'} \quad.\nonumber
\end{eqnarray}
In terms of the matrices $\vec{\alpha}$ and $\beta$, defined in (\ref{eqn:BetaAlpha1}) we get
\begin{eqnarray}
\S=\frac{1}{\sqrt{1+\frac{\g u}{\c^2}}}\,\beta\, \mathrm{exp}\left({\frac{\vec{\g}\cdot\vec{\alpha}\, t}{2c}}\right)\quad. \label{eqn:A8}
\end{eqnarray}
where $\vec{\g}=(0,0,\g)$.
The action (\ref{eqn:A03}) now reads
\begin{eqnarray}
S[\bar\psi',\psi']=\int\bar\psi'(x^{\nu'}) \hspace{0.4 \linewidth} & &\\ \times \,\left[\i \hbar \gamma^{\mu'}(u)\left(\p_{\mu'}+\S^{-1}\p_{\mu'}\S\right)-m\c\right]\psi'(x^{\nu'}) & &\nonumber\\
\times \,\, \left(1+\frac{\g u}{\c^2}\right)\,\d {x'}^4\quad. \nonumber & &
\end{eqnarray}
Finally the additional term $\gamma^{\mu'}(u)\S^{-1}\p_{\mu'}\S=0$ turns out to be zero and, thus, Eq.~(\ref{eqn:A8}) results in Eq.~(\ref{eqn:DiracAction1})
%

\section{Reminder on Covariant Derivatives} \label{section:B}
What follows is a short reminder on covariant derivatives, needed to formulate the covariant Maxwell equations in Rindler spacetime. For a detailed discussion of this topic, see for instance \cite{Carrol}, \cite{Wald}.
The covariant derivative $\nabla_\mu$, acting on a tensor $A^\nu$ with one upper or $A_\nu$ with one lower index, is connected to the partial derivative by
\begin{eqnarray}
\nabla_\mu A^\nu=\p_\mu A^\nu +\Gamma^\nu_{\mu\rho} A^\rho\quad,\\
\nabla_\mu A_\nu=\p_\mu A_\nu -\Gamma^\rho_{\mu\nu} A_\rho\quad,
\end{eqnarray}
where the Christoffel symbols $\Gamma^\rho_{\mu\nu}$ are constructed by first partial derivatives of the metric
\begin{eqnarray}
\Gamma^\rho_{\mu\nu}=\frac{1}{2} g^{\rho\sigma}\left(\p_\mu g_{\sigma \nu}+\p_\nu g_{\sigma \mu}-\p_\sigma g_{\mu\nu}\right)\quad.
\end{eqnarray}
Therefore the covariant Lorentz gauge condition for a four-vector potential $A_\mu=(\Phi/c,-\vec{A})$ is
\begin{eqnarray}
\nabla_\mu A^\mu=\p_\mu A^\mu +\Gamma^\mu_{\mu\rho} A^\rho=0\quad,
\end{eqnarray}
where the indices of $A^\mu=g^{\mu\nu} A_\nu$ are risen up with the inverse metric $g^{\mu\nu}$. The same way the electromagnetic field strength tensor with two upper indices  $F^{\mu\nu}=g^{\mu\rho}g^{\nu\sigma}F_{\rho\sigma}$ is constructed from $F_{\mu\nu}=\p_{\mu} A_{\nu} - \p_{\nu} A_{\mu}$. The covariant derivative of this quantity is 
\begin{eqnarray}
\nabla_\rho F^{\mu\nu}&=&\p_\rho F^{\mu\nu} +\Gamma^\nu_{\rho\sigma} F^{\mu\sigma}+\Gamma^\mu_{\rho\sigma} F^{\sigma\nu}\quad,\\
\nabla_\nu F^{\mu\nu}&=&\p_\nu F^{\mu\nu} +\Gamma^\nu_{\nu\sigma} F^{\mu\sigma} \quad, \label{eqn:AppMaxwell}
\end{eqnarray}
where in the special case of the vacuum Maxwell equations $\nabla_\nu F^{\mu\nu}=0$, for $\rho=\nu$ in (\ref{eqn:AppMaxwell}) the term $\Gamma^\mu_{\sigma\nu} F^{\nu\sigma}=0$ vanishes, which gives Maxwell equations in the form of Eq.~(\ref{eqn:Maxwell}).
%

\section{Towards geonium in Newtonian gravity} \label{section:C}
In this Appendix we discuss the steps and transformations, leading to the Hamiltonian (\ref{eqn:hatH0}), its energies $E_{n,k,\ell,s}$ and the corresponding eigenfunctions $\phi_{k,n,\ell,s}$.
Therefore, the starting point is $H_0$, from (\ref{eqn:H0}), which is the Hamiltonian of geonium, affected by a homogeneous gravitational field:
\begin{eqnarray}
H_0&=&m\c^2+ m \vec{\g}\cdot\vec{r}'+e\,\vec{r}'\cdot(\hat{Q}\cdot\vec{r}')\label{eqn:C1} \\
& & \qquad\qquad +\frac{1}{2m} \vec{\pi}^2-\frac{e\gs }{2m} \vec{B}^{(0)}\cdot\vec{s}\,, \nonumber
\end{eqnarray}
%

\emph{(i)} In the first step, we rotate the coordinate system from the laboratory frame with the coordinates $\vec{r}'=(x',y',z')$, where the $z'$-axis points into direction of $\vec{g}$, to the frame of trap geometry with $\vec{r}=(x,y,z)$, where the $z$-axis and $\vec{B}^{(0)}$ are aligned:
\begin{eqnarray}
x'=x \cos\varbeta- z\sin\varbeta \,,\,y'=y\,,\,z'=z \cos\varbeta+ x\sin\varbeta. \qquad
\end{eqnarray}
This transformation introduces the angle $\theta$ between the direction of acceleration and the magnetic field, as it is shown in FIG.~\ref{Fig2}. The Hamiltonian (\ref{eqn:C1}) in the rotated system reads
\begin{eqnarray}
H_0=m\c^2+ m \g (z\cos\varbeta+ x\sin\varbeta)-\frac{eV}{4L^2}(x^2+y^2-2z^2)\quad\nonumber & & \\
  +\frac{1}{2m} \vec{p}^2 + \frac{e^2 B^2 }{8m}(x^2+y^2) -\frac{e B }{2m} \,L_3-\frac{e\gs B }{2m} \, s_3\,.\quad\qquad\label{eqn:C3}& &
\end{eqnarray}
%
%

\emph{(ii)} In the second step, the coordinate dependent Newtonian potential in (\ref{eqn:C3}) is absorbed by an additional coordinate transformation to coordinates $\vec{R}=(X,Y,Z)$, that shifts the coordinate center by a constant vector, such that it coincides with the new equilibrium position of the electron motion, as it is shown in FIG.~\ref{Fig2}: 
\begin{eqnarray}
x=X + 2\g \sin\varbeta/\omega_z^2 \,\,,\,\, y=Y\,\,,\,\, z=Z-\g \cos\varbeta/\omega_z^2\,, \quad\quad
\end{eqnarray}
 Moreover, we have to absorb the upcoming constant shift of the linear momentum in $Y$-direction by an unitary transformation $U(Y)=\mathrm{exp}\left(\i m\g \sin\varbeta\, \omega_c Y/(\hbar\omega_z^2)\right)$ of the Hamiltonian $\tilde H_0=U(Y)^{\dagger}H_0U(Y)$ and the electron wave function $\phi(\vec{R})=U(Y)\phi_0(\vec{R})$. After that, the coordinate dependent Newtonian potential in  $\tilde H_0$ is replaced by an additive constant value:
\begin{eqnarray}
\tilde H_0&=& m\c^2+ \frac{m}{2}\omega_z^2Z^2 +\frac{m}{8}(\omega^2_c-2\omega_{z}^2)(X^2+Y^2) \qquad \label{eqn:HGeo}\\
& & +\frac{1}{2m} \vec{P}^2 -\frac{\omega_c}{2}(L_3+\gs\, s_3)+\frac{m\g^2}{\omega_z^2}\left(1-\frac{3}{2}\cos^2\varbeta\right) \,,\nonumber
\end{eqnarray}
where we replaced the electromagnetic quantities by the frequencies (\ref{eqn:cyclotron}) and (\ref{eqn:axial}). Expressing $\vec{R}$ in cylindrical coordinates 
\begin{eqnarray}
X&=&\rho\,\cos\varphi \quad,\quad Y=\rho\,\sin\varphi\quad,\quad Z=\zeta \,,
\end{eqnarray}
we obtain the Hamiltonian $\tilde H_0$ as shown in (\ref{eqn:hatH0}).
The solution $\phi^{k,n,\ell,s}(\vec{R})=U(Y)\phi^{k,n,\ell,s}_0(\vec{R})$ to the eigenvalue problem (\ref{eqn:eigenvalueeq}) is:
\begin{eqnarray}
& &\phi^{k,n,\ell,s}(\vec{R})=\nonumber\\[0.5em]
& & \frac{1}{\sqrt{2\pi}}U(\rho\sin\varphi)\e^{\i  (\ell-n) \varphi}
R^{n,\ell}(\rho)W^k(\zeta)\,|s\rangle\quad, \label{eqn:wavefunctions}
\end{eqnarray}
where
\begin{eqnarray}
U(\rho\sin\varphi)&=&\mathrm{exp}\left(\frac{\i m\g \sin\varbeta\, \omega_c \,\rho\sin\varphi}{\hbar\omega_z^2}\right)\\
R^{n,\ell}(\rho)&=&\left(\frac{m\omega_{\bar c}}{2\hbar}\right)^{(1+\ell-n)/2}\sqrt{\frac{2\,n!}{\ell!}}\label{eqn:Radial}\\
& &\qquad \times \e^{-\frac{m\omega_{\bar c}\rho^2}{4\hbar} }\rho^{\ell-n}\mathcal{L}^{\ell-n}_{n}\left(\frac{m\omega_{\bar c}}{2\hbar}\rho^2\right)\nonumber\\
W^k(\zeta)&=& \left(\frac{m\omega_z}{\pi\hbar}\right)^{1/4}\frac{2^{-k/2}}{\sqrt{k!}}\\
& & \qquad \times \, \e^{-\frac{m\omega_z\zeta^2}{2\hbar} }\mathcal{H}_k\left(\sqrt{\frac{m\omega_z}{\hbar}}\zeta\right)\nonumber
\end{eqnarray}
where we defined 
$\omega_{\bar c}=\sqrt{\omega^2_c-2\omega_{z}^2}$ and have used the Laguerre polynomials $\mathcal{L}^{\ell-n}_{n}$, the Hermite polynomials $\mathcal{H}_n$ and the spin basis $|s\rangle=|\pm1/2\rangle$. The wave functions (\ref{eqn:wavefunctions}) are normalized with respect to the scalar product $\langle\phi_1|\phi_2\rangle_{0}\equiv \int \phi_1^\dagger\phi_2 \,\rho\, \d\rho\d \varphi\d \zeta$.
%

We are allowed to apply first order perturbation theory as usual in flat space, using the scalar product $\langle\phi_1|\phi_2\rangle_{0}$, since the result would coincide with the energy corrections, obtained from scalar product (\ref{eqn:ScalarProduct}) to desired order. The justification of this procedure is motivated very detailed in \cite{Parker80}.

\end{document}